\documentclass[12pt]{article}
\usepackage{graphicx}
\usepackage{amssymb}
\usepackage[english]{babel}
\usepackage{makeidx,verbatim}
\parskip=\medskipamount
\newcommand{\be}{\begin{equation}}
\newcommand{\ei}{\end{equation}}
\begin{document}

\begin{center}{\bf\Large From Copenhagen to neo-Copenhagen interpretation}


{\bf \large Willem M. de Muynck\\
Theoretical Physics, Eindhoven University of Technology,
Eindhoven, the Netherlands}
\end{center}
\begin{abstract}
Positive and negative features of the Copenhagen interpretation
are discussed. As positive features can be mentioned its
pragmatism and its awareness of the crucial role of measurement.
However, the main part of the contribution is devoted to the
negative features, to wit, its pragmatism (once again), its
confounding of preparation and measurement, its \textit{classical}
account of measurement, its completeness claims, the ambiguity of
its notion of correspondence, its confused notion of
complementarity. It is demonstrated how confusions and paradoxes
stemming from the negative features of the Copenhagen
interpretation can be dealt with in an amended interpretation, to
be referred to as `neo-Copenhagen interpretation', in which the
role of the measuring instrument is taken seriously by recognizing
the quantum mechanical character of its interaction with the
microscopic object. The ensuing necessity of extending the notion
of a quantum mechanical observable from the Hermitian operator of
the standard formalism to the positive operator-valued measure of
a generalized formalism is demonstrated to yield a sound
mathematical basis for a transition from the Copenhagen
contextualistic-realist interpretation to the neo-Copenhagen
empiricist one. Applications to the uncertainty relations and to
the Bell inequalities are briefly discussed.
\end{abstract}



\section{Introduction}\label{sec1}
Interpretations of physical theories are neither true nor false.
Even if they are not completely internally consistent they may be
thought to provide useful rules of correspondence between the
mathematical formalism and the physical reality a theory purports
to describe. It is well-known that the Copenhagen interpretation
of quantum mechanics is no exception. Even 80 years after its
conception it is impossible to say that there exists a unique and
internally consistent interpretation by that name. On the
contrary, this interpretation has been characterized (Feyerabend
\cite{Feyerabend68}) as ``not a single idea but a mixed bag of
interesting conjectures, dogmatic declarations, and philosophical
absurdities.''

In my contribution to a `conference devoted to 80 years of
Copenhagen Interpretation' it therefore is not possible to present
a eulogy. On the contrary, my main attention will be directed
toward the many imperfections of the Copenhagen interpretation.
Among these I do not count the pragmatic attitude the founding
fathers of quantum mechanics have displayed while developing the
theory so as to be able to come to grips with the ``strange''
results their experiments confronted them with. On the contrary,
at that time a pragmatic approach, not bothering too much about
the physical meaning of the applied mathematics, turned out to be
advantageous to rapid scientific progress. However, maintaining
such a pragmatism for 80 years may have become detrimental both
with respect to development of fundamental insights as well as
experimental applications. As seen from table \ref{table1} the
list of negative features is considerably longer than that of the
Copenhagen interpretation's positive features.
\begin{table}[h]
\begin{tabular}{|l|l|}
\cline{1-1}\cline{2-2}
 {\bf Positive features}   & {\bf Negative features} \\
\cline{1-1}\cline{2-2}
    +1. pragmatism     & -1. pragmatism \\
    +2. crucial role of measurement & -2. confusion of preparation
    and measurement\\
 & -3. classical account of measurement\\
 &  -4. completeness claims\\
  & -5. ambiguous notion of correspondence\\
   & -6. confused notion of complementarity  \\
\hline
\end{tabular}
\caption{Positive and negative features of the Copenhagen
interpretation} \label{table1}
\end{table}
Thus, we have to deal with the Copenhagen preoccupation with
measurement, having as a consequence a confusion of preparation
and measurement. Another negative feature is the
\textit{classical} account of measurement, overlooking a crucial
property of measurements performed on \textit{microscopic} objects
by means of \textit{macroscopic} measuring instruments. Moreover,
each of the three characteristics of the Copenhagen
interpretation, viz. completeness, correspondence, and
complementarity, is liable to criticism and will be criticized.

Any of these issues might be sufficient to reject the Copenhagen
interpretation as a useful interpretation of quantum mechanics.
Yet, there is one important positive feature in the Copenhagen
appreciation of quantum mechanics as a description of microscopic
reality, viz. its recognition of the \textit{crucial role played
by measurement}, which probably outweighs its imperfections. This
feature, often considered a weakness of the interpretation, is
actually an asset, making quantum mechanics different from
classical theory in a truly revolutionary way. It makes quantum
mechanics a paradigm of the structuralist methodology (e.g. Suppe
\cite{Sup77}) in which the Einsteinian idea of a physical theory
as a `description of an \textit{objective} reality (i.e. being
independent of the observer, including his measuring instruments)'
has been relinquished.

\section{The Copenhagen interpretation and measurement}\label{sec2}

\subsection{Crucial role of measurement}\label{sec2.1}

Classical mechanics is generally presented as representing
knowledge about an \textit{objective} reality, i.e., a reality as
it is independently of being observed by human observers, and, in
particular, not being interfered with by their measuring
instruments. The impossibility of an interpretation of quantum
mechanics as an \textit{objective} description of reality was a
main reason for Einstein to disqualify quantum mechanics as not
being an adequate physical theory. On the other hand, Bohr was
deeply convinced that the existence of a non-vanishing quantum of
action $h\neq 0$ entailed a fundamental impossibility of such
objective knowledge: only knowledge could be obtained about
`microscopic reality as it is \textit{in interaction with a
measuring instrument}', i.e., \textit{contextual} knowledge (cf.
section~\textbf{4}). The Heisenberg uncertainty relations were
considered as evidence of this, the quantized interaction of an
object with different measuring instruments (of incompatible
observables) being held responsible for the impossibility of
obtaining simultaneously sharp knowledge about e.g. position and
momentum.

According to Bohr and Heisenberg quantum mechanics did not
represent any objective knowledge, and it did not need to do so
because such knowledge would not be verifiable. It should
nevertheless be stressed here that the discussion between Bohr and
Einstein was not over the question of verifiability, but over the
question of whether it is possible to obtain knowledge about a
microscopic object \textit{without in any way interacting with it}
\cite{EPR}, i.e. whether quantum mechanics can yield an
\textit{objective} description (Einstein), or whether we should
content ourselves with a \textit{contextual} one (Bohr). Only
after many years we have gradually become convinced (e.g., by the
Kochen-Specker \cite{KS} and the Bell \cite{Bell64} theorems) that
it is \textit{im}possible to attribute sharp values to quantum
mechanical observables as \textit{objective} properties possessed
by a microscopic object independently of any measurement. Bohr's
contextualism, to the effect that quantum mechanical observables
are well-defined only within the context of a measurement serving
to measure that very observable, can be seen as an indication that
physical theories do not have the absolute property of being
either universally true or universally false, but that they are
applicable on a certain domain of experimentation, for quantum
mechanics its domain of validity being co-determined by the
measurement arrangement. The Copenhagen insight that measurement
plays an ineradicable role in the interpretation of quantum
mechanics is a fundamental epistemological attainment, in a more
general form finding application far beyond quantum mechanics and
even physics.

\subsection{Confusion of preparation and measurement}\label{sec2.2}
\begin{figure}
\leavevmode \centerline{
  \includegraphics[height=1.5in]{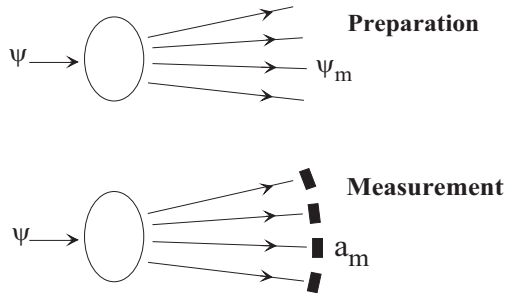}
  }\caption{Preparation and measurement.}
  \label{fig1}
\end{figure}

However, the Copenhagen preoccupation with measurement has its
drawback. As a consequence of the emphasis put on `measurement',
there was a tendency to formulate all physical operations in terms
of this particular one. Measurement was seen where there is none.
Thus, in Heisenberg's version of the Copenhagen interpretation a
measurement result actually corresponds to a property of the
microscopic object in its \textit{final} state (Heisenberg
\cite{Heis27}). Consider, for instance, the Stern-Gerlach
experiment. Here, an atom with non-vanishing angular momentum,
after traversing an inhomogeneous magnetic field may finally be
found in one of a number of different beams (cf.
figure~\ref{fig1}).
Finding the atom in a beam is interpreted as a determination of
the value of a component of the atom's angular momentum. It is
important to note here that for performing such a measurement it
is essential that detectors are set up in the beams. Without these
detectors the experimental arrangement does not correspond to a
measurement but to a conditional \textit{preparation} (e.g. de
Muynck \cite{dM2002}, section~3.2.6 and 3.3.4), allowing to
perform a measurement of an \textit{arbitrary} observable
conditional on the beam (state $|\psi_m\rangle$) the atom happens
to be in. Unfortunately, following Heisenberg \cite{Heis27} the
Stern-Gerlach arrangement is generally interpreted as a
measurement also if no detectors are present (sometimes referred
to as a `preparative measurement'), thus allowing the experiment
to satisfy (be it only approximately) the von Neumann projection
postulate, stating that the conditionally prepared state
$|\psi_m\rangle$ equals the eigenstate $|a_m\rangle$ corresponding
to the eigenvalue that allegedly would have been found if the
measurement had been performed by inserting a detector. It is
rather evident that if von Neumann's postulate holds if the
detector is absent, its presence will in general change the state
appreciably, thus showing the absurdity of the projection
postulate as a measurement principle.

\begin{figure}[b]
\leavevmode \centerline{
\includegraphics[height=3.7cm]{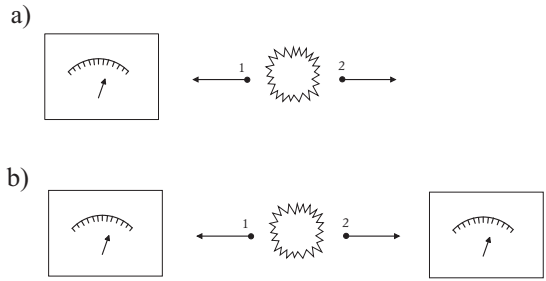}
}\caption{a) EPR experiment; b) EPR-Bell experiment.}\label{fig2}
\end{figure}

As an example of the confusion of preparation and measurement
should be mentioned here the EPR setup (figure~\ref{fig2}a),
presented in \cite{EPR} as a \textit{measurement} of a property of
particle $2$ (without in any way interacting with that particle).
Unfortunately, Bohr accepted the EPR proposal as such, even though
it is not really a measurement of particle $2$ but rather a
preparation of that particle. The EPR setup should be compared
with an EPR-Bell experiment (figure~\ref{fig2}b), like the ones
performed by Aspect and co-workers \cite{Aspect} to test the Bell
inequalities by means of a joint measurement on \textit{both}
particles. Unfortunately, experiments of the EPR-Bell type are
generally referred to as EPR experiments, thus veiling the
fundamental difference of these experiments.

\subsection{Bohr's classical account of measurement}\label{sec2.3}

The stress laid by Bohr on the macroscopicity of measuring
instruments, and the necessity to account for the results of
measurement in \textit{classical} terms (cf. section~\textbf{4})
is well-known and need not be discussed here. It should be
emphasized, however, that by thus restricting the attention
exclusively to the macroscopic part of the measurement it was
neglected that a measurement performed on a microscopic object
must also have a \textit{microscopic} part, viz. a part that is
sensitive to the \textit{microscopic} information, so as to be
able to realize information transfer from the microscopic object
to the measuring instrument. For instance, in the Stern-Gerlach
experiment the initial phase of the measurement (nowadays
generally called the `pre-measurement') in which the atom
interacts with a magnetic field so as to slightly influence the
atom's position, is a microscopic process which should be
described quantum mechanically rather than classically. In the
amplification phase of the measurement the atom's position can be
thought to change classically so as to create a sufficient
distance between the beams to distinguish one from the other. It
therefore seems that the microscopic rather than the macroscopic
part of the measurement is the interesting one. Bohr's exclusive
attention to the macroscopic part has not contributed to achieving
this insight, to say the least. In particular, the development of
the generalized formalism is a consequence of a rigorous
application of quantum mechanics to the interaction of the
microscopic object and the measuring instrument (cf.
section~\textbf{3.2}).

It should be noted that von Neumann and Heisenberg did consider a
quantum mechanical description of measurement. However, they did
not challenge Bohr's classical description; their purpose was only
to prove that a quantum mechanical description of measurement is
compatible with Bohr's classical one. In particular, reliance on
the projection postulate c.q. restriction to measurements
approximately satisfying it, was responsible for the fruitlessness
of these by itself praiseworthy undertakings.

\section{(In)completeness of quantum mechanics}\label{sec3}
\subsection{(In)completeness in a wider sense}\label{sec3.1}
 Two senses of
`(in)completeness of quantum mechanics' should be distinguished,
viz. `(in)completeness in a \textit{wider} sense', dealing with
the question of `whether hidden variables are possible', and
`(in)completeness in a \textit{restricted} sense', addressing the
question of `whether the standard formalism of quantum mechanics
describes all possible measurements within the domain of quantum
mechanics'. In the past the main interest has been attracted by
the first issue. I will not go into this, but want to restrict
myself to demonstrating the relative futility of the discussion by
comparing the answer given by Copenhagen physicists A.D. 1935
(viz. ``Quantum mechanics is complete; there are no hidden
variables'') to the answer most practicing physicists would
probably give A.D. 2007 (viz. ``Quantum mechanics is incomplete;
`no go' theorems of hidden variables theories were found to be
defective, possibly at the expense of introducing nonlocality
(Bohm, Bell)'').

Contrary to the Copenhagen idea that the wave function yields a
complete description of an individual object, it seems nowadays to
be generally recognized that the wave function does rather
describe an \textit{ensemble}. This insight has been gained by
careful analysis of interference experiments at low incident
particle rates in which impacts of \textit{individual} particles
can be registered. From these experiments it can be inferred that
what is described by the wave function (i.e. the interference
pattern) is not generated by an individual particle but by an
ensemble of such particles. The Copenhagen idea that the wave
function would be a complete description of an \textit{individual}
object does seem to be obsolete by now.

Does this mean that Einstein was right when attempting in the EPR
paper \cite{EPR} to prove the incompleteness of quantum mechanics?
Not completely so. Actually, Einstein was bound to fail to do so
because he attempted to prove the existence of hidden variables by
identifying these with `values of quantum mechanical observables'.
The impossibility of this was rigorously demonstrated, only much
later, by Kochen and Specker's proof \cite{KS} of the theorem
named after them. However, this does not exclude the possibility
of subquantum theories involving hidden variables of a different
kind. In any case does it seem to be evident that within any
theory aspiring at a description of `reality behind the quantum
mechanical phenomena' the contextuality, noticed for the first
time by Bohr and rejected by Einstein, will have to be taken into
account.

\subsection{(In)completeness in a restricted sense}\label{sec3.2}
\begin{figure}[t]
\leavevmode \centerline{
\includegraphics[height=3.3cm]{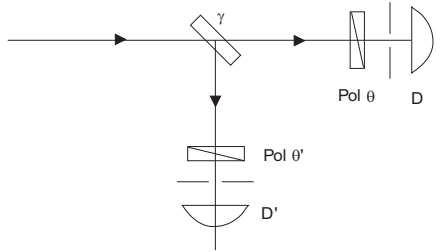}
}\caption{Which-way polarization measurement of a
photon.}\label{fig4}
\end{figure}

Whereas at this moment the question of `(in)completeness in a
wider sense' is not experimentally relevant because we do not have
any indication which kind of experiments should be performed so as
to transcend the quantum mechanical description in an
observationally relevant way, the situation is different as
regards `(in)completeness in a restricted sense', at least if we
restrict ourselves to the standard (textbook) formalism of that
theory (in which measurement probabilities of quantum mechanical
observables are represented by expectation values of the
projection operators of the spectral representations of Hermitian
operators). The standard formalism is easily seen to be
`incomplete in a restricted sense'. For instance, consider the
`which-way polarization measurement of a photon' depicted in
figure~\ref{fig4}. Here a photon has probability $\gamma$ to be
transmitted by a nonpolarizing semi-transparant mirror toward a
polarization measurement setup in direction $\theta$, and
probability $1-\gamma$ to be reflected toward a polarization
measurement setup in direction $\theta'$.
Since the detection probabilities of detectors $D$ and $D'$ are
given by $p_D =\gamma \langle E^\theta_+\rangle$ and $p_{D'}
=(1-\gamma) \langle E^{\theta'}_+\rangle$, respectively,
$E^\theta_+$ and $E^{\theta'}_+$ being projection operators, it is
evident that the detection probabilities are not described by the
standard formalism (e.g. $\gamma E^\theta_+ \neq (\gamma
E^\theta_+)^2$). From the whole range $0\leq \gamma\leq 1$ of
possibilities of the present experiment only the `set of measure
zero $\gamma =(0,1)$' satisfies the standard formalism.

This example can be supplemented by many other ones; actually,
there are very few realistic experiments satisfying the standard
formalism (see e.g. de Muynck \cite{dM2002}). If it is taken into
account that the interaction of object and measuring instrument is
a \textit{quantum mechanical} process (cf. section~\textbf{2.3}),
detection probabilities are seen to be expectation values of the
operators $M_m$ of a \textit{positive operator-valued measure
(POVM)} corresponding to the resolution of the identity $\{M_m\},
\; M_m\geq O,\; \sum_m M_m =I$, rather than a projection-valued
measure (PVM) corresponding to the orthogonal decomposition
$\{E_m\}, \; E_m^2=E_m,\; \sum_m E_m =I$ of the standard
formalism. Indeed, denoting by $\rho_o$ and $\rho_a$ the initial
density operators of microscopic object and measuring instrument,
respectively, the final density operator is given by $\rho_{oaf} =
U \rho_o \rho_a U^\dagger,\;U = e^{-\frac{i}{\hbar}HT}$, $T$ the
interaction time. Then detection probabilities of the pointer
positions (represented by a pointer observable having spectral
representation $\{E_{am}\}$) are given by
\[p_m = Tr_{oa} \rho_{aof} E_{am} = Tr \rho_o M_m,\; M_m = Tr_{a}
\rho_a U^\dagger E_{am}U.\]
 From the physical properties of the distribution $\{p_m\}$ it
follows that the operators $M_m$ satisfy all properties of the
elements of a POVM defined above. It should be noted that there is
no single reason to require idempotency of these operators.

\section{The correspondence principle (mature form)}\label{sec4}

Disregarding here Bohr's early use of the correspondence principle
(in the sense of requiring a classical limit for defining the
notion of a quantum mechanical observable), I want to restrict
myself here to his later characterization, according to which the
following requirements should be met:
\begin{enumerate}
 \item A quantum mechanical observable is exclusively defined
within the context of the measurement serving to measure that
observable.
 \item Experimental arrangement and measurement results
have to be described in classical terms.
\end{enumerate}
Usually these requirements are considered part of the
complementarity principle, while restricting the notion of
`correspondence' to the requirement of a classical limit. It,
however, seems more appropriate to stick to Bohr's later use, and
restrict `complementarity' to considerations on \textit{pairs} of
`complementary observables' only (cf. section~\textbf{5}).

The importance of the first item has been stressed already in
section~\textbf{2.1}, whereas the second item was criticized in
section~\textbf{2.3}. It is my intention in the present section to
demonstrate that by not consistently sticking to the first item
Bohr introduced an ambiguity in his notion of `correspondence',
thus being responsible for much confusion. The cause of this
ambiguity can be traced back to not distinguishing between two
different possibilities of what is meant by the word `phenomenon',
viz. either `a macroscopically observable property of a
\textit{microscopic} object', or rather `a property of a
\textit{macroscopic} measuring instrument obtained by letting the
instrument interact with the object (e.g. a pointer triggered by
the microscopic object to take a certain position on a measurement
scale)'. From the way Bohr reacted to the EPR proposal \cite{EPR}
it is evident that he had in mind the first possibility. Indeed,
since no measuring instrument is present for particle $2$ (cf.
figure~\ref{fig2}a) Bohr could accept Einstein's proposal to view
the EPR measurement setup as a \textit{measurement} on particle
$2$ only if the measurement result would be taken as a property of
that particle.

As a consequence of this Bohr applied his correspondence principle
to EPR in an inconsistent way. As a matter of fact, Bohr did not
recognize the correlation of two quantum mechanical observables
$A_1$ and $A_2$, proposed in the EPR paper as a \textit{classical}
correlation ($A_i$ either position or momentum of particle
$i,\;i=1,2$), as an ordinary \textit{quantum mechanical}
observable. He granted Einstein well-definedness of correlation
observable $A_1A_2$ although only $A_1$ is measured in the EPR
experiment. But, according to his correspondence principle the
correlation observable $A_1A_2$ is well-defined only in an
EPR-Bell experiment. Hence, Bohr's conclusion that Einstein's
`element of physical reality' is ambiguous (due to its dependence
on the context of the measurement of either $P_1$ or $Q_1$) must
be completed by the observation that Bohr's conclusion is itself a
consequence of an ambiguity as regards the meaning of a quantum
mechanical observable. Actually, like Einstein, also Bohr never
considered the possibility of looking upon a quantum mechanical
measurement result as a pointer position rather than as a property
of the microscopic object. Only by noting the difference between
EPR experiments and experiments of the EPR-Bell type (cf.
figure~\ref{fig2}) the inconsistency of Bohr's reaction to EPR
could become obvious.

Unfortunately, negligence of the difference between EPR and
EPR-Bell experiments has been perpetuated until the present day.
As a consequence it is still widely thought that properties of
particle $2$ may be nonlocally influenced by the measurement
performed on particle $1$. It will be seen in section~\textbf{6}
that there is no reason for such a conclusion if quantum
mechanical measurement results are assumed to correspond to
pointer positions of a measuring instrument.

\begin{figure}[t]
\leavevmode \centerline{
\includegraphics[width=10cm]{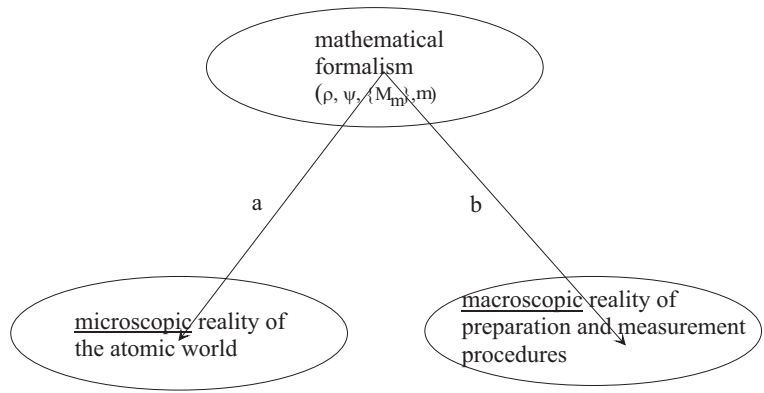}
}\caption{\em\small Realist (a) and empiricist (b) interpretations
of the mathematical formalism of quantum mechanics.} \label{fig6}
\end{figure}

In order to evade confusions like the one discussed here it is
necessary to be as precise as possible with respect to the
correspondence between the mathematical formalism of quantum
mechanics and physical reality. For this reason two different
notions of `correspondence' should be distinguished here, viz.
(restricting ourselves to standard observables)
\begin{enumerate}
\item Realist correspondence:\\
    Quantum mechanical observable $A$, as well as its values $a_m$ refer to properties of the microscopic
    object.
  \item Empiricist correspondence:\\
    Quantum mechanical observable $A$ is a label of a measurement procedure, $a_m$ is a label of a pointer
    position.
\end{enumerate}
More generally, we should distinguish two different
interpretations of the mathematical formalism of quantum
mechanics, viz. a \textit{realist} interpretation (either
objectivistic or contextualistic) interpretation (cf.
figure~\ref{fig6}(a)) in which quantum mechanics is thought to
describe microscopic reality most in the same way classical
mechanics is generally thought to describe macroscopic reality,
and an \textit{empiricist} interpretation (cf.
figure~\ref{fig6}(b)) in which state vector and density operator
are thought to correspond to preparation procedures, and quantum
mechanical observables (either standard or generalized) correspond
to measurement procedures and the phenomena induced by a
microscopic object in the macroscopically observable pointer of a
measuring instrument.

The empiricist interpretation is particularly suited to encompass
the generalization from the standard notion of a quantum
mechanical observable to the generalized one referred to in
section~\textbf{3.2}. An important application of the empiricist
interpretation to the \textit{generalized} formalism is that POVMs
can often be interpreted as labeling \textit{nonideal} measurement
procedures, not registering `reality as it objectively is', but
taking into account possible disturbing influences introduced by
the measurement itself. Thus, of two measurement procedures
labeled by POVMs $\{M_m\}$ and $\{N_n\}$, respectively, such that
 \begin{equation}
M_m =\sum_n \lambda_{mn} N_n,\; \lambda_{mn}\geq 0, \sum_m
\lambda_{mn}=1, \label{4.1}
 \end{equation}
the first is a nonideal version of the second, the nonideality
matrix $(\lambda_{mn})$ representing the conditional probability
that a measurement of POVM $\{M_m\}$ yields result $m$ if a
measurement of POVM $\{N_n\}$ would have given result $n$. In the
following sections applications of this will be discussed.
Restricting ourselves to discrete spectra it is possible to
quantify the nonideality by means of certain properties of the
nonideality matrix. A convenient quantity is the average row
entropy
\begin{equation}\label{4.2}
   J_{(\lambda)} = - \frac{1}{N} \sum_{mn} \lambda_{mn} \ln \frac{\lambda_{mn}}
{\sum_{n'} \lambda_{mn'}}.
\end{equation}

\section{The complementarity principle}\label{sec5}
The complementarity principle must be seen as a `no go' theorem as
regards general applicability of `correspondence as obtains in
classical physics': it was soon established that the notion of
`correspondence' as defined in section~\textbf{4} cannot be
applied simultaneously to quantum mechanical position and momentum
observables, or, more generally, to \textit{incompatible} standard
observables (corresponding to non-commuting Hermitian operators).
From discussions of the so-called `thought experiments' it was
seen that measurements of incompatible standard observables had
mutually exclusive measurement arrangements, in such a way that
measurement of one observable would disturb incompatible ones. The
mutual disturbance of measurement results of standard observables,
when measured simultaneously, was thought to be described by the
Heisenberg uncertainty relation
 \begin{equation}
 \Delta A\Delta B\geq \frac{1}{2}|\langle
 \psi|[A,B]_-|\psi\rangle|
 \label{5.1}
 \end{equation}
 derived from the standard formalism.

It is important here to remind a criticism by Ballentine
\cite{Bal70}, to the effect that the Heisenberg uncertainty
relation (\ref{5.1}) does not refer to `joint measurement' because
it can be tested by means of \textit{separate} ideal measurements
of observables $A$ and $B$. Hence, it is impossible that this
inequality is an expression of mutual disturbance as found in the
`thought experiments'; it should be seen as a restriction of our
\textit{preparing} abilities rather than of our \textit{measuring}
ones. Probably Einstein was closer to the real meaning of the
Heisenberg uncertainty relation (viz., as a restriction on the
values of standard deviations in an \textit{ensemble}) than were
Bohr and Heisenberg, who seem to have been jumping to conclusions
by associating the results of their considerations on the `thought
experiments' with equation (\ref{5.1}).

Perhaps Bohr and Heisenberg should not be blamed too much for this
error because they did not have at their disposal the generalized
formalism referred to in section~\textbf{3.2}, which turns out to
be necessary to clarify the situation. It, indeed, is possible by
means of the POVMs of the generalized formalism to prove that in
trying to jointly measure incompatible standard observables mutual
disturbance necessarily occurs (Martens, de Muynck \cite{MadM90};
also de Muynck \cite{dM2002}, section~7.10), thus corroborating
the results obtained by Bohr and Heisenberg by studying `thought
experiments'.

\begin{figure}[t] \leavevmode \centerline{
\includegraphics[height=6cm]{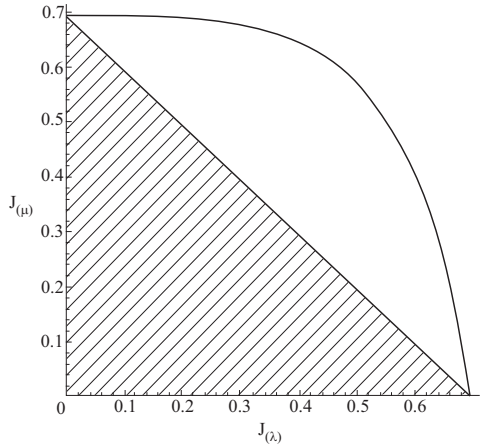}
} \caption{Parametric plot of $J_{(\protect\lambda) }$ versus
$J_{(\protect\mu) }$ as a function of $\gamma$. The shaded area is
forbidden by the Martens inequality
\protect{(\ref{5.4})}.}\label{fig7}
\end{figure}

In order to do so a bivariate POVM $\{R_{mn}\}$ was considered,
the expectation values of which describing the joint probability
distributions of the two pointers intended to yield the
measurement results of the incompatible observables $A=\sum_m a_m
E_m$ and $B=\sum_n b_n F_n$ to be measured jointly. Then these
latter probability distributions are given by the expectation
values of the marginals $\{\sum_nR_{mn}\}$ and $\{\sum_mR_{mn}\}$,
respectively. We will call a measurement procedure a joint
measurement of $A$ and $B$ if $ \sum_n R_{mn} = E_m$ and $ \sum_m
R_{mn} = F_n$. It is not difficult to prove that this is possible
only if $A$ and $B$ are compatible. However, if  $A$ and $B$ are
incompatible it is possible that the marginals of $\{R_{mn}\}$
correspond to \textit{nonideal} measurements (cf. (\ref{4.1})) of
$A$ and $B$, respectively. Thus, the POVM $\{R_{mn}\}$ describes a
joint \textit{nonideal} measurement of standard observables $A$
and $B$ if nonideality matrices $(\lambda_{mm'})$ and
$(\mu_{nn'})$ exist such that
\begin{equation}\label{5.2}
\begin{array}{l}
\sum_n R_{mn} =\sum_{m'} \lambda_{mm'} E_{m'},\; \lambda_{mm'}\geq
0,\;\sum_m \lambda_{mm'}=1,\\
\sum_m R_{mn} =\sum_{n'} \mu_{nn'} F_{n'},\; \mu_{nn'}\geq
0,\;\sum_n \mu_{nn'}=1.
\end{array}
\end{equation}
It is easily seen that measurement procedures exist satisfying
(\ref{5.2}). Thus, the measurement arrangement depicted in
figure~\ref{fig4} is a joint nonideal measurement of incompatible
standard polarization observables in directions $\theta$ and
$\theta'$, respectively. Indicating a detection event by $+$, and
non-detection by $-$, we find
\begin{equation}\label{5.3}
(R_{mn})= \left(
\begin{array}{cc}
O & \gamma  E^\theta_+\\
(1-\gamma) E^{\theta'}_+ &  I-\gamma E^\theta_+ - (1 - \gamma)
 E^{\theta'}_+
\end{array} \right),
\end{equation}
event $`++'$ having probability zero, and $I-\gamma E^\theta_+ -
(1 - \gamma) E^{\theta'}_+$ corresponding to the possibility that
the photon is not detected by either one of the detectors $D$ or
$D'$ but is absorbed in one of the analyzers. It is
straightforward to calculate the nonideality matrices for this
experiment as
\begin{equation}\label{5.8}(\lambda_{mm'}) =
\left(\begin{array}{cc} \gamma & 0\\1-\gamma&
1\end{array}\right),\; (\mu_{nn'})=\left(\begin{array}{cc}
1-\gamma & 0\\\gamma& 1\end{array}\right).\end{equation}
 From (\ref{5.8}) it is easily seen how the mutual disturbance of the
measurement results of the two standard observables changes if the
parameter $\gamma$ is changed. In particular, complementary
behaviour is evident, deviation of one nonideality matrix from the
unit matrix being maximal whenever the other's is minimal. This
can be expressed by means of the nonideality measure (\ref{4.2}).
For measurements of the type described by (\ref{5.2}) it is
possible (Martens, de Muynck \cite{MadM90}) to derive the
following general inequality,
\begin{equation}\label{5.4}
    J_{(\lambda)} + J_{(\mu)} \geq -\ln \{\max_{mn} Tr
E_m F_n\},
\end{equation}
to be referred to as the Martens inequality. It is important to
note that, contrary to the Heisenberg inequality (\ref{5.1}), the
Martens inequality does not depend on the state vector or density
operator but in an unambiguous way is a property of the
measurement procedure only. From an application of (\ref{5.4}) to
the joint nonideal measurement of polarization observables
depicted in figure~\ref{fig7} it is seen that the Martens
inequality
yields a perfect illustration of the results obtained by Bohr and
Heisenberg when studying the `thought experiments'. Indeed, their
physical observations based on these experiments were completely
correct; only their reluctance to exert a full-blown
\textit{quantum mechanical} analysis of the measurement process
prevented them from a sufficiently complete analysis.

It, incidentally, should be noticed that the mutual disturbance
considered here does refer to the final pointer positions of the
measuring instruments. This should be distinguished from `mutual
disturbance in a preparative sense' as discussed in the Copenhagen
literature, referring to the final state of the microscopic object
rather than to the final state of the measuring instrument.
Actually, complementarity can manifest itself in two different
ways, viz., as a property of `preparation' described by the
Heisenberg inequality, and as a property of `joint measurement'
described by the Martens inequality. To a large extent the
Copenhagen confusion with respect to `complementarity' has its
origin in the confusion with respect to preparation and
measurement noted in section~\textbf{2.2}.

\section{The Bell inequalities and nonlocality}\label{sec6}
As is well-known the Bell inequalities are satisfied if all
standard observables that are involved are mutually compatible.
This entails the conclusion that \textit{in}compatibility is a
necessary condition for the Bell inequalities to be violated. But,
due to the `postulate of local commutativity' (stating that
observables measured in causally disconnected regions of
space-time must be compatible) incompatibility of observables is a
local affair. Hence, violation of the Bell inequalities must have
a \textit{local} origin. It is the intention of the present
section to demonstrate that, indeed, violation of the Bell
inequalities may be caused by \textit{local} mutual disturbance in
joint nonideal measurements of incompatible standard observables
rather than by nonlocal influences.

\begin{figure}[t]
\leavevmode \centerline{
\includegraphics[height=4.5cm]{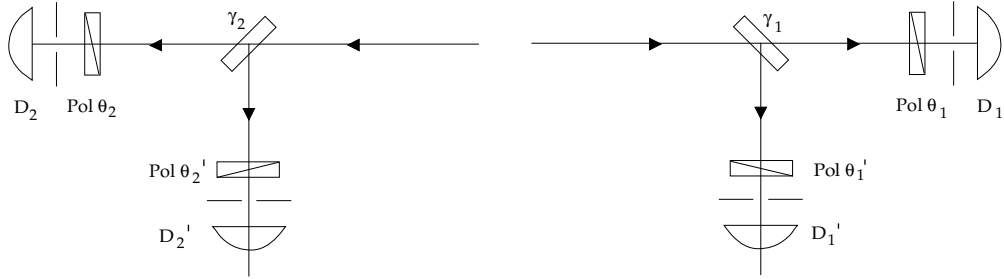}
} \caption{\em \small Generalized EPR-Bell experiment.}
\label{fig8}
\end{figure}

In order to do so we consider the generalized EPR-Bell experiment
depicted in figure~\ref{fig8}, the difference with the Aspect
experiments \cite{Aspect} being that the mirrors indicated by
$\gamma_1$ and $\gamma_2$ are `semi-transparant, and stationary'
rather than `completely reflecting, and either stationary or
switching'.
We consider here two measurements of the type depicted in
figure~\ref{fig4}, simultaneously performed on particles $1$ and
$2$. The Aspect experiments are special cases of the present
experiment, satisfying $(\gamma_1,\gamma_2) =(1,1), (1,0), (0,1)$
and $(0,0)$,
 respectively. For general $(\gamma_1,\gamma_2)$ the (quadrivariate) POVM
$\{R_{m_1n_1m_2n_2}\}$ follows directly from the bivariate POVM
(\ref{5.3}) as
\begin{equation}\label{6.1}
R^{(\gamma_1\gamma_2)}_{m_1n_1m_2n_2} =
R^{(\gamma_1)}_{m_1n_1}R^{(\gamma_2)}_{m_2n_2},\end{equation}
 in which the parameters $\gamma_1$ and $\gamma_2$ are added to
distinguish the two arms of the interferometer. The expectation
value of POVM (\ref{6.1}) yields the quadrivariate probability
distribution of a joint nonideal measurement of four standard
polarization observables in directions $\theta_1, \theta'_1,
\theta_2$, and $\theta'_2,$ respectively. There is mutual
disturbance of the type discussed in section~\textbf{5}
\textit{separately} in each of the arms of the interferometer. As
is to be expected from the postulate of local commutativity there
is no mutual disturbance between different arms of the
interferometer.

Let us finally see where the violation of the Bell inequalities in
the Aspect experiments \cite{Aspect} comes from (see de Muynck
\cite{dM2002}, chapter~9). In order to do so it is important to
note that these experiments can be demonstrated to violate the
Bell inequalities only by combining results obtained within four
\textit{different} arrangements for measuring bivariate
probabilities. By contrast, the bivariate probabilities derived
from the generalized experiment for fixed $(\gamma_1, \gamma_2)$
from the experimentally obtained quadrivariate one do satisfy the
Bell inequalities because they are obtained within one single
measurement setup. This can be traced back to the fact that each
individual particle pair yields a quadruple $(m_1,n_1,m_2,n_2)$ of
measurement results (pointer readings); this is a sufficient
condition for the statistics to have a Kolmogorovian character.
This evidently holds true for each of the Aspect experiments
separately, but \textit{not} for measurement results of the four
experiments pasted together.

Violation of the Bell inequalities is evidence of the
non-Kolmogorovian character of the statistics. Indeed, by changing
the measurement arrangement in the experiments performed by Aspect
the quadruples are changed in such a way that an \textit{octuple}
of measurement results of the four measurements corresponding to
$(\gamma_1,\gamma_2) =(1,1), (1,0), (0,1)$, and $(0,0)$, cannot be
accommodated into one single quadruple. The cause of this is seen
to be a changing of the mutual disturbance in one arm of the
interferometer by changing the measurement arrangement in that arm
(and analogously in the other arm). These changes are completely
accounted for by the the idea of complementarity in a joint
measurement of incompatible standard observables as discussed in
section~\textbf{5}. Nonlocal influences between different arms are
not necessary to explain the ensuing violation of the Bell
inequalities.

It should finally be stressed that the latter conclusion, although
derived from (generalized) quantum mechanics, is actually a
conclusion about the \textit{physical processes} obtaining in
experiments of the kind discussed here. Therefore, the often-heard
assertion that quantum mechanical locality might be compatible
with subquantum nonlocality (the latter, allegedly, being
necessary to cope with derivations of the Bell inequalities from
local hidden variables theories), is not applicable: it is utterly
unreasonable to suppose that one and the same phenomenon would
have completely different \textit{physical} explanations within
different theories. If violation of the Bell inequalities has a
\textit{local} explanation within a quantum mechanical description
of certain phenomena, it should have a \textit{local} explanation
within any theory describing the reality behind these phenomena.
Rather than acquiescing in an unfortunate marriage of `observable
locality' with `unobservable nonlocality' by relying on such fancy
expressions like `entanglement', `inseparability' or `passion at a
distance', it seems more appropriate to perform as precise as
possible analyses of measurement processes, and to study the
relations between its measurement results and the properties of
microscopic objects as described in different theories.

\section{From Copenhagen to neo-Copenhagen
interpretation (summary)}\label{sec7}

Due to Bohr's reference to `phenomena' the Copenhagen
interpretation has an empiricist reputation (e.g. Reichenbach
\cite{Reich44}). However, on closer scrutiny Bohr's `phenomenon'
is probably too much inspired by the rather primitive detection
methods (like flashes on a scintillation screen) of that time. It
does not take into account more modern measurement procedures in
which the final observational data are processed so as to be only
vaguely reminiscent of their microscopic origin: what is seen by
the observer is often just a graph coming out of his printer.
Contrary to the Copenhagen interpretation (which, for the main
part, is a realist interpretation, be it of a contextualistic
kind) the neo-Copenhagen interpretation is an empiricist one,
adapted to the generalized formalism of quantum mechanics
necessary to also describe more modern experiments (cf. de Muynck
\cite{dM2002}, chapter~8).

Comparing, in summary, the neo-Copenhagen interpretation with the
Copenhagen one with respect to the notions of completeness,
correspondence and complementarity, we arrive at the following
results:
 \begin{table}[h]
\begin{tabular}{lll}

\underline{interpretation}:& \underline{Copenhagen}& \underline{neo-Copenhagen}\\
\underline{completeness}: &&\\
 in a wider sense:&yes (claimed) &no\\
 in a restricted sense (standard qm): &yes (claimed)&no\\
 in a restricted sense (generalized qm): && yes\\
 \underline{correspondence}:  & contextualistic-realist&empiricist\\
 &partly classical&completely qm\\
  \underline{complementarity}: &&\\
  preparation:&&Heisenberg inequality \\
  joint measurement:&Heisenberg inequality&Martens inequality
\end{tabular}
\end{table}

 Notwithstanding the considerable differences does it nevertheless seem
appropriate to pay a tribute to Copenhagen by referring to that
name in the new interpretation, because in the latter the
important development away from the classical idea of `physical
theory as an \textit{objective} description of reality', which was
started by the Copenhagen interpretation, has been continued, and
has been pushed to a logical completion.

Acknowledgement:

The author would like to thank the Faculty of Applied Physics of
Eindhoven University of Technology, and in particular professor
Thijs Michels, for providing the facilities that made this work
possible.






\end{document}